\documentstyle{article}
\begin{document}
\begin{center}
{\bf CPP-GIANT MAGNETORESISTANCE AND THERMO-ELECTRIC POWER OF
MULTILAYERS}\\[7pt]
S.~Krompiewski\footnote{corresponding author} \\[5pt]
Institute of Molecular Physics, P A S \\
Smoluchowskiego 17, Pl-60-179 Pozna\'n, Poland \\[5pt]
and U.~Krey \\[5pt]
Institut f\"ur Physik II, Universit\"at Regensburg, D-93040
Regensburg, Germany
\end{center}
\begin{abstract}
Oscillations of 
magnetoresistance and thermo-electric power $(TEP)$ {\it vs.}
 both nonmagnetic
spacer as well as ferromagnetic slab thicknesses are studied in the
current-perpendicular-to-plane $(CPP)$ geometry, 
in terms of the single-band tight-binding model.
 The spin-dependent conductance
has been calculated from the Kubo formula by means of a recursion
 Green's
function technique, and the $ TEP $ directly 
from the well-known Onsager relations.
 In general, the observed oscillations
may have either just one or two periods. 
In the latter case
the long period of oscillations, related to spectacular beats,
 is apparently of
$ non-RKKY $ type. 
The relative $ TEP $ oscillations  are strongly
enhanced in comparison with those of the giant magnetoresistance, have
the same periods, but  different phases and a negative bias. 
\end{abstract} 
\section{Introduction} 

Since the discovery of giant magneto-resistance ({\it GMR}) in magnetic
multilayers \cite{l:ba,l:bi} there has been a great deal of interest
in studying this unusual phenomenon also by theoretical methods [3 --7].
The
{\it GMR} was first interpreted phenomenologically with emphasis put on
the role of interface roughness \cite{l:ca},
 afterwards quantum mechanical
methods were applied, too ({\it e.g.}~\cite{l:le}). There is no doubt
 now that the {\it GMR} does appear even in systems without
 any imperfections, since
 this was clearly demonstrated by {\it ab initio} computations 
in \cite{l:sche}. 
In the present paper we study the current-perpendicular-to-plane 
{\it(CPP)-GMR} effect in layered systems 
of the type $ W/F/S/F/W$, where {\it W} stands for a semi-infite ideal
lead wire, $ F$ for a ferromagnet, and {\it S} for the non-magnetic
spacer. 

\section{Method and Results}

The calculation technique, we have
developed, is based on the Green's function recursion method
\cite{l:lee,l:as}. After having performed the Fourier transformation
in the {\it x-y} plane, the equation of motion of the
Green's function has been reduced (for s.c. structure) to a simple
tridiagonal eigen-problem quite easy to cope with.
The conductance of electrons of spin $\sigma$ is
given by the Kubo formula, which in turn may be expressed
in terms of Green's functions as follows, \cite{l:lee,l:as}:

\begin{equation} \label{eqn3} 
\Gamma_{\sigma} = \frac{8 e^2 N_x N_y}
{h\,(2\pi)^2} \int_{BZ} d^2{k}_{\|} \left[ G_{\sigma}''(i,i)
G_{\sigma}''(i-1,i-1) - G_{\sigma}''(i,i-1)^2\right]\,.
\end{equation}

Here the index $i$ labels the $z$-planes, $G''$
stands for the imaginary part of $G$,
 $e^2/h$ is the conductance quantum,
$N_x N_y$ the cros-ssection area, and the integration is over the
2-dimensional Brillouin zone. 
Hereafter, both the hopping and the lattice
constant are taken as energy- and length-units. 
In addition to the conductance we have also calculated the
 thermo-electric power $S$,
 which is related to $\Gamma$ by the following well-known
formula:

 \begin{equation} \label{eqn6}
S_\sigma = -\pi^2 k_B^2 [T /(3 |e|)]
{\rm (d/dE)} \log \Gamma_\sigma (E). \end{equation}
Analogously to the {\it GMR},
we define the
corresponding ''giant magneto-thermo-electric power'' by

\begin{equation} \label{eqn7}
GMTEP= (S^{\uparrow\uparrow}_{\uparrow} +
 S^{\uparrow\uparrow}_{\downarrow})/
 (S^{\uparrow\downarrow}_{\uparrow} + 
 S^{\uparrow\downarrow}_{\downarrow})
-1\,\,,
\end{equation}
where the superscripts refer to the parallel and
antiparallel magnetization configurations of the two ferromagnets,
whereas the  subscripts refer to the carrier spin.\\
It can be seen from Fig.~1 
that the {\it CPP-GMR} oscillates as a function of the thickness
$n_f$ of {\it the ferromagnetic slab} with a short period of
$\approx 2$ monolayers (ML).
Additionally, in Fig.~1 {\it pronounced beats} are
seen with typical repetition lengths of roughly $10$ ML. 
 In case of Fig.~2 with 
the spacer thickness
($n_s$) being varied, one gets an intermediate period of $\sim 4.5$
ML, without any beats.
It is clear from these Figs. that the oscillations of the {\it GMTEP}
are even
stronger than those of the {\it GMR}, although they have a 
negative bias and a different phase than those of the {\it GMR}.
In any case the {\it GMR} oscillations
decay always roughly as $n_f^{-1}$.\\
It was shown in \cite{l:ma} that the oscillatory behaviour of $\Gamma$ 
may arise both from wavenumbers with stationary $k_z(k_x,k_y;E_f)$,
as well as from particular cut-off wavenumbers ($k_x,k_y$),
for which a spectral density vanishes abruptly at one of the
ferromagnet/spacer--interfaces.\\ 
One can roughly estimate the ''RKKY-type'' wave numbers
$k_z$ from the asymptotic equation (valid actually for large
thicknesses and constant potentials) 
\begin{equation}\label{eqkz}
k_z(k_x,k_y;E_f)=\arccos [(V_\sigma -E_f)/2-\cos k_x -\cos k_y]\,.
\end{equation}
By finding the extremum with respect to $k_x, k_y$, with
$E_F$=2.5, we get $k_z^{(1)}=0.722$ for $V_\sigma =0$ (for all 
the carriers within the spacer and minority-spin electrons in the 
ferromagnets), and $k_z^{(2)}=1.721$ for $V_{\uparrow}=-1.8$
(the majority-spin carriers in the ferromagnets).
The corresponding wavelengths can then be
calculated as in \cite{l:al} from the expression $\lambda(p,q) =
\pi/(p k_z^{(1)}+q k_z^{(2)})$, which yields $\lambda(0,1)$ = 1.825
($V$ = -1.8) , i.e. roughly the short period of two
monolayers (2 ML) seen in Fig.~1. It is interesting that in
Fig.~2, where only the spacer thickness is varied, only a longer
wavelength of $\sim 4.5$ ML is visible, which would correspond to
$\lambda(1,0) = 4.35$.\\  
In order to get the beats seen in Fig.~1 one would need, apart from
the wave characterized by $k_z^{(2)}$, another wave with a wavenumber 
very close to it. As there is no way to get such a wavenumber
from Eq.~(\ref{eqkz}), we suppose it to be of {\it non-RKKY} character.
More insight into the nature of such {\it non-RKKY} oscillations 
can only be gained by taking into account localized and resonance 
states in addition to extended ones used for the sketchy estimations 
above. Although the localized states do not contribute directly to
 the conductance, certainly they modify 
the density of states and thereby its value at $E_F$.
Those unextended states are easily seen in our method
as singularities (spread over a large range of energy)
 of the local densities of states -- we are going to study 
them analytically in the nearest future. 

In conclusion, within an s-band tight-binding
model combined with the Green's function recursion method, we have
reduced the calculation of the {\it CPP-GMR} effectively to one
dimension. We have computed the {\it CPP-GMR} for perfect systems
composed of two ferromagnets separated by a non-magnetic
spacer and sandwiched between ideal infinite lead wires. Unusual beats
of the {\it GMR vs.}~the thickness of the ferromagnetic slab 
thickness have been found and suggested to be due to $non-RKKY-type$ 
oscillations.  Similar results, with even stronger beats, have also
been obtained for the ''giant magneto-thermo-electric power''. 


{\bf Acknowledgements}

This work has been carried out under 
the bilateral project DFG/PAN 436 POL. We also thank the  Pozna\'n,
Munich and Regensburg Computer Centres for computing time.

\newpage
\input epsf
%
%
%
\epsfxsize=11.5cm \epsfbox{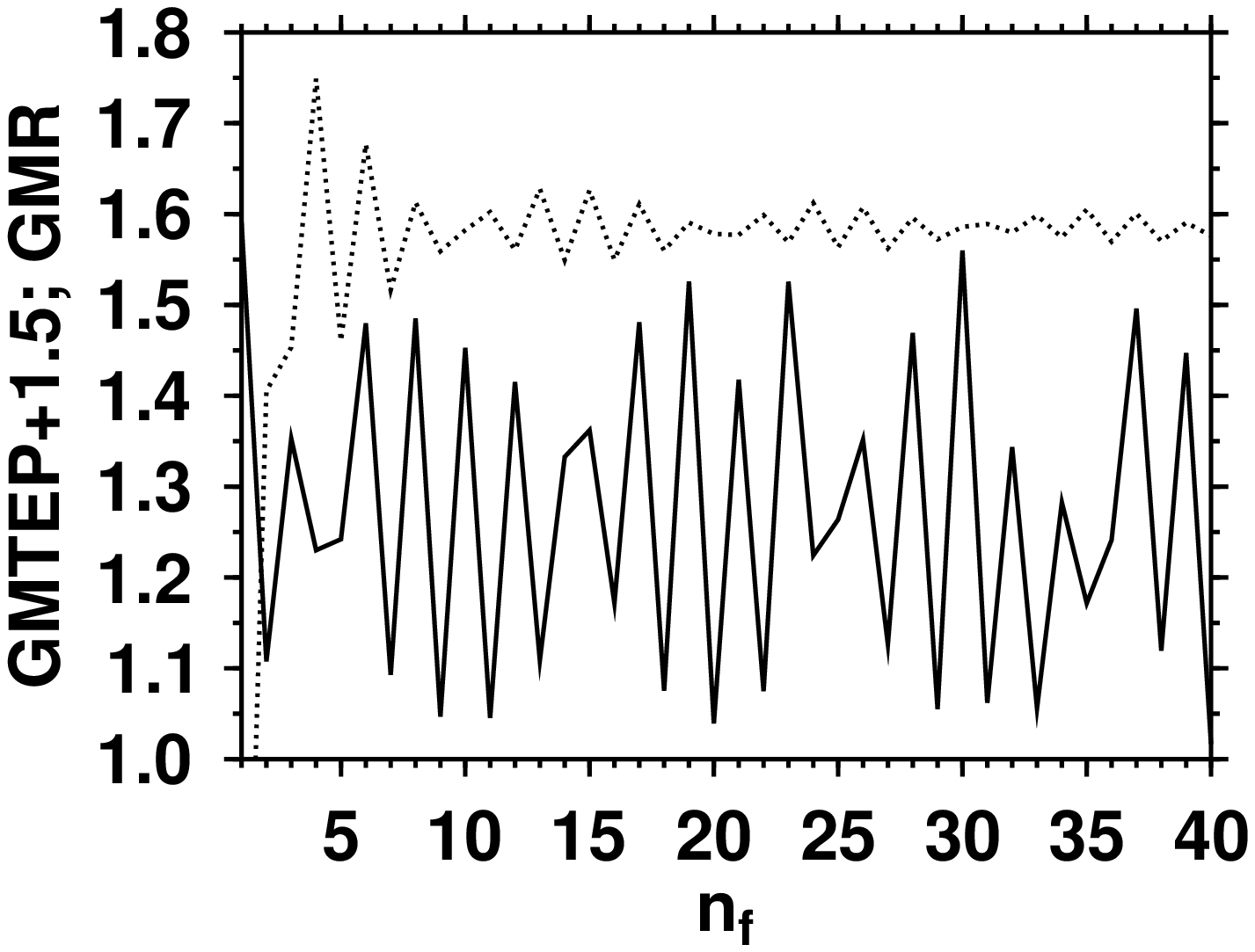}

 \underbar{Fig.~1}: 
{\it CPP-GMR} (dotted) and 
{\it GMTEP} of the system $ n_fF/n_sS/n_fF$ 
with $n_s$ = 5 
(where $ F $ and $ S $ stand for ferromagnet and spacer, respectively)
sandwiched between two semi-infinite ideal lead wires. 
Majority spin electrons have the potential
$V_{\uparrow}= -1.8$ in the ferromagnet (all
other potentials are 0), $ E_F$ = 2.5.

\epsfxsize=11.5cm
\epsfbox{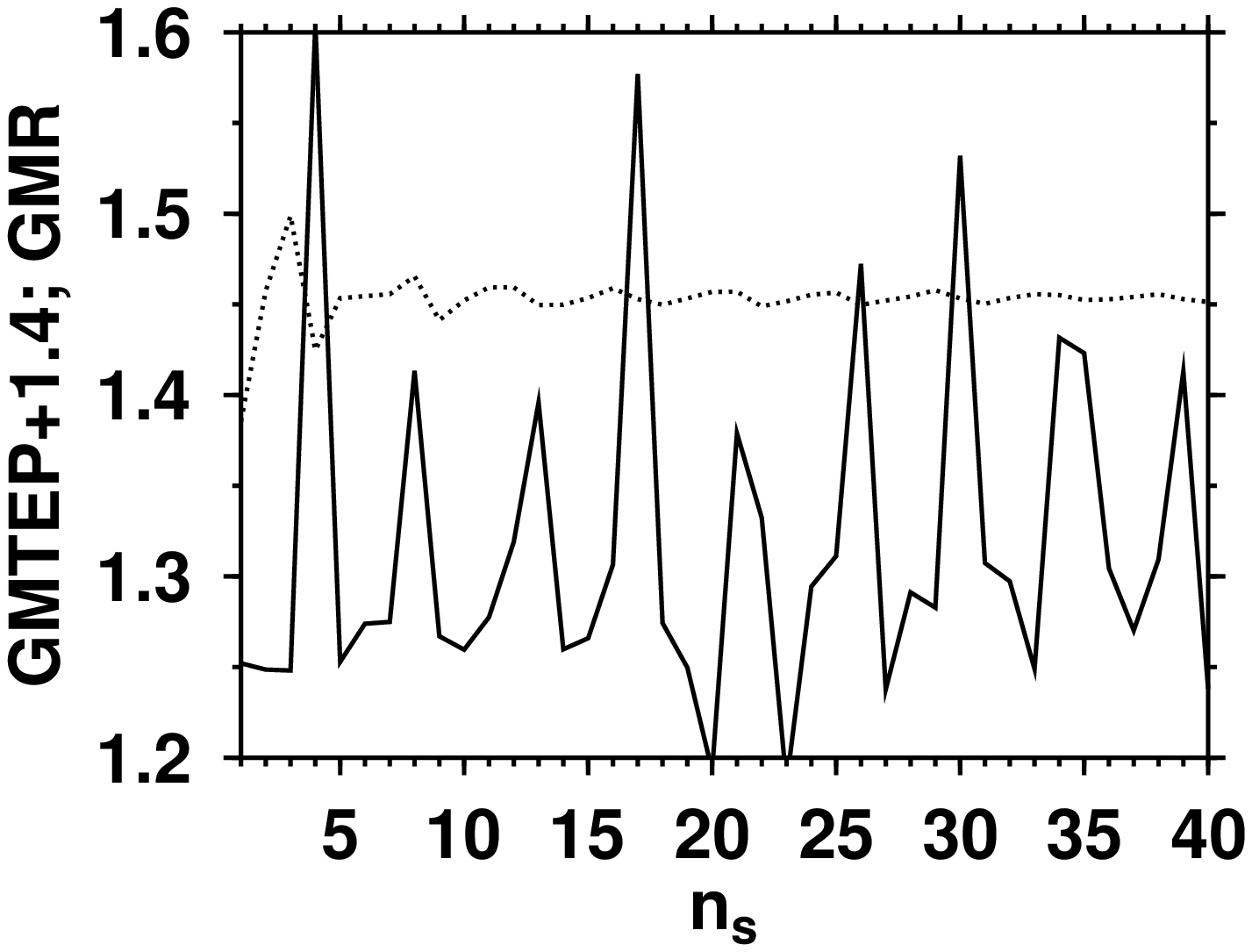}
\underbar{Fig.~2}: 
 Same as Fig.~1, but with $n_f$ = 3 on both sides, as a function
of the spacer thickness $n_s$.

\begin{thebibliography}{99}
\bibitem{l:ba} M.N. Baibich, J.M. Broto, A. Fert, F.N. Van Dau,
 F. Petroff, P. Etienne, G. Creuzet, A.Friedel, J. Chazelas,
 {\it Phys. Rev. Lett.} {\bf 61}, 2472 (1988).
\bibitem{l:bi} G. Binasch, P. Gr\"unberg, W.Zinn, {\it Phys. Rev. B}
 {\bf 39}, 4828 (1989).
\bibitem{l:ca} R.E. Camley, J. Barna\'s, {\it Phys. Rev. Lett.} 
{\bf 63}, 664 (1989).
\bibitem{l:le} P.M. Levy, S. Zhang, A. Fert,
               {\it Phys. Phys. Lett.} {\bf 65}, 1643 (1990).
\bibitem{l:sche} K.M. Schep, P.J. Kelly, G.E.W. Bauer,
{\it Phys. Rev. Lett.} {\bf 74}, 586 (1995).
\bibitem{l:as} Y. Asano, A. Oguri, S. Maekawa, {\it Phys. Rev. B }
{\bf 48}, 6192 (1993).
\bibitem{l:ma} J. Mathon, M. Villeret, H. Itoh, {\it Phys. Rev. B}
 {\bf 52}, R69 (1995).
\bibitem{l:lee} P.A. Lee, D.S. Fisher,
{\it Phys. Rev. Lett.} {\bf 47}, 882 (1981).
\bibitem{l:al} J. d'Albuquerque e Castro, J. Mathon, M. Villeret,
 D.M. Edwards, {\it Phys. Rev. B} {\bf 51}, 12 876 (1995).
\end{thebibliography}
\end{document}